**Energy flow in biological system: Bioenergy transduction of V$_1$-ATPase molecular rotary motor from *E. hirae***

**Ichiro Yamato**[a,b,*], **Takeshi Murata**[b,c], **Andrei Khrennikov**[d]

[a]Department of Biological Science and Technology, Tokyo University of Science, 2641 Yamazaki, Noda-shi, Chiba 278-8510, Japan, [b]Department of Chemistry, Graduate School of Science, Chiba University, 1-33 Yayoi-cho, Inage, Chiba 263-8522, Japan, [c]JST, PRESTO, 1-33 Yayoi-cho, Inage, Chiba 263-8522, Japan, and [d]International Center for Mathematical Modelling in Physics and Cognitive Sciences, Linnaeus University, Växjö, SE-351 95, Sweden

[*]**Corresponding Author:**

Ichiro YAMATO

Department of Chemistry, Graduate School of Science, Chiba University, 1-33 Yayoi-cho, Inage, Chiba 263-8522, Japan

TEL/FAX: +81-43-290-2794

E-mail: iyamato@rs.noda.tus.ac.jp

**Abstract**

We classify research fields in biology into those on flows of materials, energy, and information. As a representative energy transducing machinery in biology, our research target, V$_1$-ATPase from a bacterium *Enterococcus hirae*, a typical molecular rotary motor is introduced. Structures of several intermediates of the rotary motor are described and the molecular mechanism of the motor converting chemical energy into mechanical energy is discussed. Comments and considerations on the information flow in biology, especially on the thermodynamic entropy in quantum physical and biological systems, are added in a separate section containing the biologist friendly presentation of this complicated question.

## 1. Classification of research fields: flows of materials, energy, and information (Asano et al., 2015b)

Buddha advocated Buddhism containing the concepts of reincarnation and causation. We, humans, developed modern natural sciences including physics, chemistry, and biology. Now we know that everything not only limiting to living organisms recycles in our universe as symbolically suggested by "reincarnation". Thus everything cannot be steady but flows changing from one form to another.

When we think of materials, it is easy to imagine such recycling, because we learn many such examples in chemistry class. Living organisms do not reincarnate by themselves but their constituents actually recycle. The concepts of energy and information are rather new after establishment of modern sciences. And they also flow changing their forms. In this note an example of typical energy flow in living organisms from chemical to mechanical energy is introduced and discussed using our recent achievement on the structural studies of V-type ATPase from a bacterium (Arai et al., 2013; Suzuki et al., 2016; Yamato et al., 2016).

Information flow in living organisms has been studied as research target of molecular biology and also as signal transduction in biochemistry. The main flow from DNA sequences to proteins has been established as central dogma. Many biological signal transduction systems, such as hormone and nervous systems, have been studied extensively. The results of such studies in molecular biology and biochemistry are used to understand our life useful especially for medical purposes.

Now that many such information flows are known, trials to understand the living systems as a whole under an unified concept of information flow have been undertaken to make birth of the establishment of quantum bioinformatics (Asano et al., 2015a, 2015b), which is supposed to be a major theme of this special issue. Two-body system interacting one another can be solved analytically but systems more than two cannot be generally solved analytically. They usually do not follow according to classical probability conservation law but they behave non-Kolmogorovian or quantum-like (Asano et al., 2016). Such complex systems thus in most cases can be ergodic, presumably producing entropy increase, which is the origin of the time arrow; see Section 3 for more detail.

Biological information transfer systems are consisted of complex networks having a large number of interacting members of molecules including DNA/RNA and proteins, organelles, cells including nervous system as in psychology, organs, individual organisms, evolution and ecosystems, societies, and even economics. Complex interacting network system such as any biological system is then understood to behave quantum-like (Asano et al., 2015a, 2015b) and is easily imagined to bring about the "causation" in our universe; everything is interconnected affecting each other, nothing or no one isolated. In the quantum formalism, such global interconnection is represented in the form of entanglement which is often treated as an exhibition of quantum nonlocality. The main distinguishing feature of the state of entanglement for a composite system $S=S_1+...+S_n$ is that any modification of the state of any subsystem $S_j$ induces modification of the "global state", the state of $S$. In this formalism, for example, expression of some concrete gene in the genome generates update of the state of the whole genome (or epigenome; see Asano et al., 2013). Therefore, everything has the meaning of its existence by the existence itself.

## 2. Bioenergy transduction of a molecular rotary motor, $V_1$-ATPase
## 2.1. Introduction

F-ATPases working as ATP synthases in mitochondria, chloroplasts, and bacterial membranes and V-ATPases working as $H^+$ pumps in eukaryotic acidic organelles and certain bacterial membranes are relatives functioning as rotary motors (Forgac, 2007; Mulkidjanian et al., 2007): Hydrolysis energy of ATP at the catalytic hydrophilic domain ($\alpha_3\beta_3$ in $F_1$ or $A_3B_3$ in $V_1$) drives the three step rotation of its axis ($\gamma$ in $F_1$ or DF in $V_1$) connected to the membrane hydrophobic rotor ring (oligomer of c in $F_o$ or $V_o$), which in turn results in the ion translocation through the interface between the rotor ring and the hydrophobic stator subunit (a in $F_o$ or $V_o$) (Fig. 1). The three dimensional structures of $F_1$ (Abrahams et al., 1994) or $V_1$ (Arai et al., 2013) have been obtained and the single-molecule analysis of the rotation revealed important facets of the rotation mechanism (Adachi et al., 2007; Minagawa et al., 2013); the correlation of the nucleotide binding with the (sub)steps of rotation (Adachi et al., 2007). However, the basic mechanism of energy transduction, chemical energy of ATP hydrolysis converted to rotational motion, has not been fully elucidated; especially how the hydrolysis of ATP resulting in the conformational change of $A_3B_3$ brings about the rotation of the DF axis.

We have been studying V-ATPase of *Enterococcus hirae*, which acts as a primary ion pump similar to eukaryotic V-ATPases but uniquely transports $Na^+$ or $Li^+$ instead of $H^+$ (Mizutani et al., 2011). The enzyme is thought to be a bacterial homologue of eukaryotic V-ATPase (Zhou et al., 2011). We have determined its crystal structures (Arai et al., 2013; Suzuki et al., 2016), finding the asymmetrical structures of $A_3B_3$ or supposedly intermediate structures during rotation driven by ATP hydrolysis.

## 2.2. Several structures of $V_1$-ATPase and $A_3B_3$ complexes showing intermediate states of the chemomechanical energy transduction

We have obtained several intermediate structures of the rotary motor as published (Arai et al., 2013; Suzuki et al., 2016). Figure 2 shows the schematic representation of the rotational conformational changes deduced from the obtained structures (Suzuki et al., 2016).

Crystal structure of $A_3B_3$ showed asymmetry without nucleotide (bound, bindable and empty forms) and that of $V_1$ showed a newly produced structure of $A_1B_1$ pair, tight form, without nucleotide (tight, bound and empty forms) depending on the binding of DF axis (Arai et al., 2013) as shown at the step (a) in the figure; the asymmetrical structure of $A_3B_3$ is supposed to be responsible for the hydrolysis order and rotation direction. $A_3B_3$ with nucleotide also showed an asymmetrical structure (bound, bound and empty forms) and $V_1$ the same as without nucleotide. Soaking of $V_1$ without nucleotide into ADP solutions gave rise new structures supposed to be intermediates during rotation/ATP hydrolysis (Suzuki et al., 2016); bound, ADP-bound and bindable-like with a low concentration of ADP, where bindable-like did not have bound nucleotide, corresponding to the step (b) in the figure, and bound, tight-like and half-closed with a high concentration of ADP, where all the forms bound the nucleotides, which corresponds to the step (c) in the figure. The structure (d) is the same as (a) after one cycle of ATP hydrolysis driving 120° rotation of the axis.

**2.3. Possible mechanism of the chemomechanical energy transduction (Suzuki et al., 2016) and future works**

The main events occurring during one ATP hydrolysis and the 120° rotation are shown in the above figure (Fig. 2) and can be summarized as follows:

Step (a): ATP at tight form is hydrolyzed (waiting for hydrolysis of ATP) producing ADP/Pi.

Step (b): Pi is released and the ADP-bound form is produced from tight form, which brings about the conformational change of the empty form to bindable-like form (waiting for ATP binding). This is the only one structure reported so far for the ATP-binding dwell in the researches of molecular rotary motors.

Step (c): ATP is bound to the bindable-like form, which may first drive the rotation of the axis or cause release of the ADP from the ADP-bound form (waiting for ADP release). Either of the two events can be sequential or concomitant, which is not yet settled by our structural studies. At present, we speculate that rotation of the axis starts before ADP-release. Concerning to the question how the axis rotates, several hypotheses have been proposed such as a typical push-pull mechanism (Kinosita et al., 2004) and a kind of thermal ratchet mechanism (Yamato et al., 2016), which was proposed by Vale and Oosawa (1990) for actomyosin system; we at present speculate it as a kind of thermal ratchet, because DF axis in $V_1$ rotates 120° in one step (Minagawa et al., 2013) and the travelling distances of interacting amino acids of the axis with the motor ring subunits during such 120° rotation in one step seem too far to exchange to the next subunits by the push-pull mechanism (Yamato et al., 2016). In other words, the chemical free energy supplied from ATP hydrolysis is utilized not to force the axis to move but to select the rotational direction of the axis driven by thermal energy producing a kind of informational entropy. In addition, it is not yet proven whether the structure (c) shown in the figure named ADP-release dwell is really the intermediate structure for this step of ATP hydrolysis. Since this structure showed bound ADP in place of ATP at the half-closed form derived from the bindable form, this structure can be the by-product structure of ADP inhibited state. When ATP instead of ADP is bound to the bindable-like form in the structure (b), it may not produce such stable structure of (c) and rather instantaneously drives the conformational change for axis rotation and ADP release to produce the structure (d). Even another possibility for the structure (c) can be considered: *E. hirae* V-ATPase has not been examined and identified as working for ATP synthesis but the structure (c) may be possibly the intermediate for ATP synthesis, reverse reaction of the ATP hydrolysis.

Step (d): Then the axis rotates to the next bound form to make it tight form, producing the original state waiting for ATP hydrolysis.

Clear from the above summary of the rotation events, the step (c) has two optional possibilities (sequential or concomitant), which are not settled yet, and moreover, the structure (c) can be the real intermediate of ATP hydrolysis/rotation or other by-product of ADP inhibition, which should be elucidated in future. Thus the first necessary and important approach is to clarify the unsettled points at this step (c) using several kinds of experimental approaches such as computer simulation, further structural studies, and biochemical/biophysical experimental approaches.

## 3. Ergodicity, Second Law of Thermodynamics and arrow of time for classical and quantum physical and biological systems

### 3.1. From ergodicity to the Second Law of Thermodynamics

We start with the basic notion of thermodynamics, *the Boltzmann entropy.* In thermodynamics, the state space $\Lambda$ consists of macroscopic configurations determined by macroscopic parameters. In the pioneer example of an ensemble of gas molecules, these are temperature, pressure and volume. Mathematically the elements of the space of macrostates $\Lambda$ are represented as sets of microstates. It is assumed that two different macrostates are represented by disjoint sets, i.e., a microstate can belong only to one of macrostates.

Denote the space of microstates by the symbol $\Omega$. In the simplest model the space of macrostates $\Lambda$ is obtained as a finite (disjoint) partition of $\Omega$: $\Lambda=\{M_1,…, M_n\}$.

(However, in the general mathematical models $\Lambda$ need not be finite or discrete.) It is assumed that the set $\Omega$ is endowed with some measure $\mu$ such that macrostates belong to the domain of definition of $\mu$.

The Boltzmann entropy of a macrostate $M$ can be defined as

$$S_B = k_B \log[\mu(M)].$$

Here $k_B$ is the Boltzmann constant. One of the basic features the Boltzmann entropy is that it is an increasing function of the measure of macrostates. Increasing of the "size" of a macrostate induces increasing of its entropy. In particular, *the most extended macrostates have the highest entropy.* (Of course, one has to be careful with the choice of the measure on the space of macrostates $\Lambda$. The above statement is true, e.g., for the Lebesgue measure on $\Lambda$, where the latter is selected as some domain in the Euclidean space, the Cartesian product of real lines.)

It is possible to show that for natural systems the Boltzmann entropy is maximal for *the equilibrium state $M_{equilibrium}$*. Thus the equilibrium state $M_{equilibrium}$ has largest value of the measure $\mu$.

Now the main problem is to provide the microstate dynamical explanation of this thermodynamic statement. Suppose that originally a system $S$, e.g., a gas molecule, was staying in some state $M_0 \in \Lambda$. Why should $S$ finally approach the domain covered by $M_{equilibrium}$? Another related question is why then $S$ stays in $M_{equilibrium}$.

The answers to these questions are given by relying to probability theory. By normalizing the measure $\mu$ one can treat it as probability: $p(M) = \mu(M)/\mu(\Omega)$. Then the probability of $M_{equilibrium}$ is the largest. From the viewpoint of probability theory, it is natural to find a system $S$ in the most probable state. Thus approaching of the equilibrium state can be considered as the dynamics (at the macrolevel) from a less probable state to a more probable state and finally to the most probable state and the latter is the equilibrium state. The presented argument is the standard explanation of the origin of the Second Law of Thermodynamics.

However, this is still a sort of the macrolevel explanation. What does happen at the microlevel? What is about the dynamics of the microstate?

To explain the origin of the Second Law of Thermodynamics in the micro-dynamical terms, we have to appeal to the very important feature of dynamical systems, *ergodicity.* As we know, an ergodic system spends in a domain *M* of its configuration space a fraction of time which is proportional to the probability of *M*. And we know that the equilibrium state is the most probable macrostate.

By using the frequency interpretation of probability (following R. von Mises) we can interpret *p(M), M $\epsilon$ Λ,* as the fraction of time an ergodic system spends in the domain *M* of *Ω* over the course of time. Now we can summarize the previous considerations in the framework of classical statistical mechanics:

1) Macrostates can be represented as subsets of the space of microscopic states *Ω*.

2) It is possible to endow *Ω* with some measure *μ* representing the natural (physically relevant) "size" of macrostates. Its normalization, *p(M)= μ(M)/μ(Ω),* can be interpreted as the probability to find a system in the macrostate *M*.

3) Theory of probability implies evolution from less probable configuration to more probable configuration and approaching with time the most probable configuration. In the entropic terms this means increasing of the Boltzmann entropy $S_B$ (the Second Law of Thermodynamics) and approaching its maximal value.

4) Ergodicity of a system implies that in the limit it will be found inside the domain $M_{equilibrium}$ practically always over the course of time. Thus $M_{equilibrium}$ is the stationary state of the dynamical system.

The increase of entropy is the basic quantitative characteristics of the irreversibility of processes in isolated systems and the asymmetry between future and past. *The increase of entropy determines the arrow of time* (in an isolated system).

**3.2. The Second Law of Thermodynamics for isolated quantum system**

For quantum systems, the straightforward generalization of the Second Law of Thermodynamics is meaningless, because, for an isolated quantum system, the von Neumann entropy,

$$S(\rho)=Tr\ [\rho\ ln\ \rho],$$

does not change in the process of unitary evolution described by the Schrödinger equation.

At the same time *all natural Schrödinger dynamics are ergodic* – in the sense of coincidences of time and ensemble averages (von Neumann, 1929, 1955). These are dynamics with Hamiltonians having nondegenerate (discrete) spectra (distinct eigenvalues).

It seems that constancy of entropy in combination with ergodicity is a sign of striking difference from the classical case. However, this difference is only an artifact of the use of von Neumann entropy $S(\rho)$ in attempting to obtain a quantum analog of the Second Law of Thermodynamics. As was pointed out by von Neumann (1929), this entropy is not appropriate for thermodynamic considerations. This entropy is about intrinsic information contained in the quantum state $\rho$. However, essential part of this information is not approachable with the aid of macroscopic measurements. Thus, instead of von Neumann entropy $S(\rho)$, there has to be considered its modification corresponding to macro-measurement of energy, see (von Neumann, 1929, 1955) for details. For this kind of quantum entropy, it is possible to prove the quantum version of the Second Law of Thermodynamics.

Its formulation contains two types of conditions: a) related to the microdynamics of a system; b) restrictions on the type of macro-measurement of energy. The first condition is the mentioned condition of non-degeneracy of the spectrum of Hamiltonian, the generator of the evolution of an isolated quantum system. The conditions on macro-measurement of energy (determining the correspondent entropy) are sufficiently technical (von Neumann, 1929, 1955). However, they are natural from the viewpoint of extraction of information about micro-system with the aid of macroscopic information devices.

In short, any natural quantum dynamics of an isolated system is ergodic and, for any meaningful procedure of macro-measurement of energy, the quantum analog of the Second Law of Thermodynamics holds true:

*Entropy based on the macro-measurement of energy increases approaching its upper bound for the state corresponding to quantum micro-canonical ensemble.*

### 3.3. The Second Law of Thermodynamics for biological systems

We now point out that all above considerations are true only for an isolated system. *For an open system, the Second Law of Thermodynamics does not hold true.*

Any biological system α is fundamentally open; it cannot survive without interaction with the surrounding environment *E*. Therefore, above considerations cannot be applied to a biological system. The Second Law of Thermodynamics is valid only for the compound system α+*E*. For example, if α is a cell, then its own entropy does not need increase. Interaction with its environment can generate decrease of entropy or make it fluctuating around a constant value. Thus (surprisingly) the order structure of a cell can be preserved only as the result of a specially designed interaction with its environment and not as the result of its isolation. Of course, conservation of entropy of α (or its fluctuations around some fixed value) can be possible only on the cost of the increase of entropy of its environment *E*.

In the idealized model α approaches a steady state. Changes in *E* induce changes in the dynamics of the state of α and hence transition to another steady state. Thus functioning of a biological system can be considered as a chain of transitions from one steady state to another.

## 3.4. The Second Law of Thermodynamics for open quantum systems

For open quantum systems, formulation and derivation of a reasonable analog of the Second Law of Thermodynamic is a complicated problem. Recently one of possible solutions was presented in a paper (Lesovik et al., 2016). It is based on the notion of *a quasi-isolated system.* This is a very special class of quantum systems interacting with their environments in a special way.

We reformulate the classical Second Law of Thermodynamics in terms of possible state-transitions. Consider the probability density of molecules of the ideal gas in some box, *f(τ, x, v)*, where *τ* is time, *x* position and *v* velocity. Then entropy is defined as

$$S = -\int dx dv f(x, v, \tau) \log(f(x, v, \tau))$$

Under the assumption that the dynamics of the probability density is driven by *the kinetic equations of classical statistical mechanics,* it is possible to prove that entropy does not decrease, i.e., that $dS/d\tau \geq 0$. This is the statement of *Boltzmann's H-theorem*. One can say that (at the classical level) in nature only state transitions are possible which prevent entropy from decreasing. (These are state transitions driven by the kinetic equations.)

Now we turn to theory of *open quantum systems*. A quantum system α is considered in interaction with its environment (bath) *E*. The dynamics of the compound system α+*E* is unitary. The compound system can be treated as an isolated system and its dynamics is described by the Schrödinger equation with Hamiltonian (the generator of evolution)

$$H = H(\alpha) + H(E) + H(\alpha, E),$$

where *H(α)* and *H(E)* are Hamiltonians of *α* and *E*, respectively, and *H(α,E)* is the interaction Hamiltonian. As was already remarked, von Neumann entropy of the compound system *α+E* is constant.

Consider now solely the dynamics of the state of the system α. It is obtained by tracing the state of the compound system with respect to the degrees of freedom of *E*. In fact, the use of such tracing is the only reasonable treatment of the problem, because the number of the degrees of freedom of *E* can be huge and it is practically impossible to model the unitary dynamics of the system + environment, *α+E*. Denote the state of α by the symbol *ρ* and the state of α+*E* by symbol *Ψ*. It is assumed that the state *Ψ* of α+*E* is a pure state, i.e., it is given by a normalized vector of a complex Hilbert state space of *α+E*. However, we cannot assume that the state of the system *α* is also a pure state. Even by starting with a pure initial state, say *ϕ*, the dynamics in the presence of interactions with an environment transfers this pure state *ϕ* into a mixed state *ρ(τ)*.[1] Therefore, in contrast to theory of isolated quantum systems, dynamics of open quantum systems cannot be represented in terms of pure states, normalized vectors belonging to complex Hilbert space. We have to proceed with density operators. We remark that any pure state can be represented by a density operator, the orthogonal projector on this state. Denote the density operator corresponding to the state *Ψ(τ)* of *α+E* at the moment *τ* by the symbol *Σ(τ)*. This is the orthogonal projector onto the vector *Ψ(τ)*.

---

[1] We recall that mathematically mixed states are represented by *density operators*, positive semidefinite Hermitian and trace one operators. A density operator is the quantum analog of classical probability distribution. The trace one condition corresponds to normalization of a probability distribution by one. If the state space of the system *α* is finite dimensional, mathematically the situation is essentially simpler. We can proceed with density matrices, without involving theory of linear operators in complex Hilbert spaces.

Then the state dynamics for the system *α* has the form *ρ(τ)= Tr'Σ(τ)*, where *Tr'* denotes the trace with respect to the degrees of freedom of the environment *E*. Consider the map, *ρ→ ρ(τ)*, the initial state *ρ* is mapped into the state at the moment of time *τ*. Denote this map by symbol *Φ(τ)*, *Φ(τ) ρ= ρ(τ)*. It can be extended, as linear map, to the space of all linear bounded operators acting in Hilbert state space of the system *α*. Such linear operators (acting in the space of linear bounded operators) are called *super-operators*. For each *τ*, the super-operator *Φ(τ)* maps density operators into density operators. Such super-operators are called *quantum operations or quantum channels*.

For an arbitrary quantum channel *Φ*, the following inequality holds (Holevo, 2010):

$$S(\Phi(\rho)) - S(\rho) \geq k_B \, Tr[\Phi(\rho) \ln \Phi(I)].$$

where *I* is the unit operator. A quantum channel is called *unital* if *Φ(I)=I*. For a unital quantum channel, the right hand side of this inequality equals to zero. Thus we obtain the open quantum systems version of the Second Law of Thermodynamics, the quantum version of Boltzmann's H-theorem: *The entropy gain during evolution is nonnegative if the system evolution can be described by the unital channel.*

It is interesting that if the state space of a system *α* is finite dimensional, then the state dynamics given by a quantum channel produces nonnegative entropy gain if and only if the channel is unital.

If a quantum channel is not unital, then quantum entropy can decrease, see the paper (Lesovik et al., 2016) for examples, and fluctuate. We remark that this is expectable, since the formulated quantum H-theorem is about open and not isolated systems. *The nontrivial quantum result is that the arrow of time corresponding irreversibility of evolution can be consistently assigned to a class of open systems. In some sense, such systems being open behave similarly to classical isolated systems.*

We point out that if the state space of *α* is infinite-dimensional then nonnegative entropy gain can be found for non-unital quantum channels. Thus, for systems with infinite dimensional state spaces, there can exist more tricky dynamics respecting the Second Law of Thermodynamics.

**3.5. Biological systems as quantum-like systems**

In series of papers (Asano et al., 2013, 2015a, 2015b, 2016) there was presented the novel approach to mathematical modeling of behavior of biological systems based on application of theory of open quantum systems. This approach is known as *quantum bioinformatics*. This is an important application of the mathematical formalism of quantum mechanics outside of physics. Quantum bioinformatics has to be distinguished from classical bioinformatics and from quantum biophysics. The latter is about study of real quantum physical processes in biological systems. In contrast to it, quantum bioinformatics explores the quantum-like ideology.

By this ideology some macroscopic biological systems, such as cells, proteins, plants, animals, humans, can exhibit behavioral patterns which match better to quantum information and probability than to their classical counterparts. Such quantum-like behavior is a consequence of contextuality of performances (e.g., gene expressions) of biological systems. They do not have ``objective'' context independent properties. They are very sensitive to context modifications. In particular, biological systems are intrinsically adaptive to their environments. The latter motivates application of theory of *open quantum systems and more general theory of quantum adaptive dynamical systems* (Asano et al., 2015a, 2015b).

Therefore, it is natural to apply to biological systems the quantum version of the H-theorem for open quantum systems. This theorem describes mathematically the class of state transitions for which the Second Law of Thermodynamics holds true, such biological processes can be endowed with arrow of time.

For the moment, the main problem is analysis of matching of the features of quantum quasi-isolated systems (Lesovik et al., 2016) and biological systems. This problem will be studied in more detail in one of our further publications.

**Acknowledgments**

We thank Dr. K. Suzuki for her help of figure drawings for us. This work was supported (A. Khrennikov) by the EU-project "Quantum Information Access and Retrieval Theory" (QUARTZ), Grant No. 721321.

**References**

Abrahams, J.P., Leslie, A.G., Lutter, R., Walker, J.E., 1994. Structure at 2.8 Å resolution of $F_1$-ATPase from bovine heart mitochondria. Nature 370, 621-628.

Adachi, K., Oiwa, K., Nishizaka, T., Furuike, S., Noji, H., Itoh, H., Yoshida, M., Kinosita, K.Jr., 2007. Coupling of rotation and catalysis in $F_1$-ATPase revealed by single-molecule imaging and manipulation. Cell 130, 309-321.

Arai, S., Saijo, S., Suzuki, K., Mizutani, K., Kakinuma, Y., Ishizuka-Katsura, Y., Ohsawa, N., Terada, T., Shirouzu, M., Yokoyama, S., Iwata, S., Yamato, I., Murata, T., 2013. Rotation mechanism of *Enterococcus hirae* $V_1$-ATPase based on asymmetric crystal structures. Nature 493, 703-707.

Asano, M., Basieva, I., Khrennikov, A., Ohya, M., Tanaka, Y., Yamato, I., 2013. A model of epigenetic evolution based on theory of open quantum systems. Systems and Synthetic Biology 7, 161-173.

Asano, M., Basieva, I., Khrennikov, A., Ohya, M., Tanaka, Y., Yamato, I., 2015a. Quantum information biology: From information interpretation of quantum mechanics to applications in molecular biology and cognitive psychology. Foundations of Physics 45, 1362-1378 (DOI 10.1007/s10701-015-9929-y).

Asano, M., Khrennikov, A., Ohya, M., Tanaka, Y., Yamato, I., 2015b. Quantum adaptivity in biology: From genetics to cognition, Springer Netherlands, Dordrecht.

Asano, M., Khrennikov, A., Ohya, M., Tanaka, Y., Yamato, I., 2016. Three body system metaphor for two slit experiment and *E.coli* lactose-glucose metabolism. Philosophical Transactions of The Royal Society A 374, 20150243 (DOI: 10.1098/rsta.2015.0243).

Forgac, M., 2007. Vacuolar ATPases: rotary proton pumps in physiology and pathophysiology. Nat. Rev. Mol. Cell Biol. 8, 917-929.

Holevo, A.S., 2010. The entropy gain of infinite-dimensional quantum channels.

Doklady Math. 82, 730–731.

Kinosita, K. Jr., Adachi, K., Itoh, H., 2004. Rotation of $F_1$-ATPase: How an ATP-driven molecular machine may work. Annu. Rev. Biophys. Biomol. Struct. 33, 245-268.

Lesovik, G.B., Lebedev, A.V., Sadovskyy, A.M., Suslov, I.V., Vinokur, V.M., 2016. *H*-theorem in quantum physics. Sci. Rep. 6, 32815.

Minagawa, Y., Ueno, H., Hara, M., Ishizuka-Katsura, Y., Ohsawa, N., Terada, T., Shirouzu, M., Yokoyama, S., Yamato, I., Muneyuki, E., Noji, H., Murata, T., Iino, R., 2013. Basic properties of rotary dynamics of *Enterococcus hirae* $V_1$-ATPase. J. Biol. Chem. 288, 32700-32707.

Mizutani, K., Yamamoto, M., Suzuki, K., Yamato, I., Kakinuma, Y., Shirouzu, M., Walker, J.E., Yokoyama, S., Iwata, S., Murata, T., 2011. Structure of the rotor ring modified with N,N-dicyclohexylcarbodiimide of the $Na^+$-transporting vacuolar ATPase. Proc. Natl Acad. Sci. USA 108, 13474-13479.

Mulkidjanian, A.Y., Makarova, K.S., Galperin, M.Y., Koonin, E.V., 2007. Inventing the dynamo machine: the evolution of the F-type and V-type ATPases. Nat. rev. Microbiol. 5, 892-899.

Suzuki, K., Mizutani, K., Maruyama, S., Shimono, K., Imai, F.L., Muneyuki, E., Kakinuma, Y., Ishizuka-Katsura, Y., Shirouzu, M., Yokoyama, S., Yamato, I., Murata, T., 2016. Crystal structures of the ATP-binding and ADP-release dwells of the $V_1$ rotary motor. Nature Comm. 7, 13235 (DOI: 10.1038/ncomms13235).

Vale, R.D., Oosawa, F., 1990. Protein motors and Maxwell's demons: does mechanochemical transduction involve a thermal ratchet? Adv. Biophys. 26, 97–134.

Von Neumann, J., 1929. Beweis des Ergodensatzes und des H-Theorems in der neuen Mechanik. Zeitschrift fur Physik 57, 30–70.

Von Neumann, J., 1955. Mathematical Foundations of Quantum Mechanics, Princeton Univ. Press, Princenton.

Yamato, I., Kakinuma, Y., Murata, T., 2016. Operating principles of rotary molecular motors: Differences between $F_1$ and $V_1$ motors. Biophys. Physicobiol. 13, 37-44.

Zhou, M., Morgner, N., Barrera, N.P., Politis, A., Isaacson, S.C., Matak-Vinković, D., Murata, T., Bernal, R.A., Stock, D., Robinson, C.V., 2011. Mass spectrometry of intact

V-type ATPases reveals bound lipids and the effects of nucleotide binding. Science 334, 380-385.

**Figure legend**

Fig. 1. A model of V-ATPase from *E. hirae*

$V_1$ indicates catalytic domain (consisting of $A_3B_3DF$) and $V_o$ indicates membrane domain (consisting of $a$-$c_{10}$). Peripheral stalk consists of E and G subunits and central stalk consists of d, D, and F subunits.

Fig. 2. Proposed model of the rotation mechanism of *E. hirae* $V_1$-ATPase

The structure models are based on the crystal structures of $2_{ATP}V_1$ (catalytic dwell; a, d), $2_{ADP}V_1$ (ATP-binding dwell; b), and $3_{ADP}V_1$ (ADP-release dwell; c) as reported (Suzuki et al., 2016). A and B subunits are shown in respective colors representing their specific conformations. Central stalk DF axis at the center is drawn in green. ATP indicated as a yellow 'P' in (a) and (d) represents an ATP molecule that is committed to hydrolysis. See text for additional details.